\documentclass[10pt]{article}

\usepackage{cite,amssymb,epsf,amsmath}

\numberwithin{equation}{section}

\def\de {\partial}
\def\cy {Calabi--Yau}
\def\cldd {Clifford(d,\,d)}
\def\cld {Clifford(d)}
\def\clss {Clifford(6,6)}
\def\hQ{\hat Q}
\def\0 {\nonumber}

\begin{document}
\setcounter{page}{0}
\begin{titlepage}
\titlepage
\rightline{hep-th/0311122}
\rightline{CPHT-RR 106.1103}
\vskip 3cm
\centerline{{ \bf \large Mirror symmetric SU(3)-structure manifolds with 
NS fluxes}}
\vskip 1.5cm
\centerline{St{\'e}phane Fidanza, Ruben Minasian
and Alessandro Tomasiello}
\begin{center}
\em Centre de Physique Th{\'e}orique, Ecole
Polytechnique\\Unit{\'e} mixte du CNRS et de l'EP, UMR
7644\\91128 Palaiseau Cedex, France\\
\end{center}
\vskip 1.5cm
\begin{abstract}

When string theory is compactified on a six-dimensional manifold with a
nontrivial NS flux turned on, mirror symmetry exchanges the flux with a
purely geometrical composite NS form associated with lack of integrability
of the complex structure on the mirror side. Considering a general class
of $T^3$-fibered geometries admitting SU(3) structure, we find an exchange
of pure spinors ($e^{iJ}$ and $\Omega$) in dual geometries under fiberwise
T--duality, and study the transformations of the NS flux and the
components of intrinsic torsion. A complementary study of action of
twisted covariant derivatives on invariant spinors allows to extend our results
to generic geometries and formulate a proposal for mirror symmetry in
compactifications with NS flux.

\end{abstract}
\vfill
\begin{flushleft}
{\today}\\
\vspace{.5cm}
\end{flushleft}
\end{titlepage}

\newpage

\section{Introduction}

Mirror symmetry is a pairing between different compactifications which
give rise to the same four--dimensional effective theory. For \cy\
compactifications it is well--understood and has played an important role,
becoming arguably the most interesting mathematical application of string
theory. More general compactifications with fluxes on manifolds which are
not Ricci--flat have become focus of much attention recently, and it would
be important to extend to these at least partially the machinery which
proved so useful for \cy s.

If we had to consider only supersymmetric vacua, our search would be
premature. The conditions on fluxes and warping to compensate
non--Ricci--flatness and preserve supersymmetry are well--understood for
some types of fluxes. To some extent, as we review later, these conditions
are even translated into mathematical requirements: the manifold has to
have SU(3) structure and fall into a certain class in the mathematical
classification of these objects.  But Bianchi identity becomes an equation
for which there is no existence theorem in the literature, unlike the
famous Yau's theorem for \cy s (not even the analogue of Calabi conjecture
seems to have been formulated: this might be a task for string theory). If
there is no singularity in the internal compact manifold, and the higher
derivative terms are not taken into account, one can actually show even
{\it non--}existence theorems.

Fortunately mirror symmetry as a more general equivalence of effective
theories, and not only of vacua, still makes sense. As emphasized in
\cite{glmw}, to have supersymmetry of the effective action SU(3) structure is
enough, without the extra requirements mentioned above, which ensure we are 
actually in a supersymmetric vacuum. Not only looking for mirror symmetric 
SU(3) manifolds makes sense, but it is sensible to expect that a formal
advance in this direction might help to understand the still elusive
problem of compactifications with fluxes. 

Much in this spirit, \cite{glmw} (building on a comment in \cite{vafa})
considered a particular case. Namely, an $H$ flux on {\sl \cy\ } manifolds
(without back--reaction: we are not dealing with a vacuum) get mapped
to so--called half--flat manifolds, a particular class of SU(3) structure 
manifolds, without any $H$ flux. The amount by which these half--flat
manifolds fail to be Ricci--flat is measured by a certain quantity called 
 {\it intrinsic torsion}. We can thus say that, in this example, $H$ flux gets
exchanged by mirror symmetry with components of the intrinsic torsion associated with lack of integrability of the complex structure. 

It is natural to wonder what happens in more general cases, when on both sides
one has both $H$ and intrinsic torsion.
(As mentioned above, this for example is necessary in order to have
supersymmetric 
vacua.) In the \cy\ case, a concrete approach to mirror symmetry is
Strominger--Yau--Zaslow (SYZ) \cite{syz}
conjecture. This states that {\it i}) every \cy\ is 
a $T^3$ special lagrangian fibration over a three--dimensional base, and
{\it ii}) mirror symmetry is T--duality along the three circles of the $T^3$.
It is natural to try and generalize this method to the present problem. 
Part {\it i}) of the conjecture came originally from considering moduli spaces
of D--branes on \cy s; generalizing this to background with fluxes {\sl does}
seem premature, and in any case we do not attempt it here, although
later we will
comment more on it. So we simply assume the manifold and flux we
start with have this property, of admitting three Killing vectors. The idea
is that the mirror transformations found in this class of examples will
generalize to some extent to the most general case. 

Having assumed this, we perform T--duality along the three isometries at once. 
T--duality will preserve four--dimensional effective theories, but since
eventually we hope this procedure could be extended to more general situations
by including singular fibers as in SYZ, we want to show why
this should be called mirror symmetry  -- for that matter,
indeed, why is there any mirror symmetry at all. A good framework for
answering this is Hitchin's method based on \clss\ spinors \cite{hitchin}. As we
review later in more detail, these are simply formal sums of forms on the
manifold.
Existence on a manifold of a \clss\ spinor without zeros
which is also {\it pure}  (annihilated by half of the gamma matrices) 
is the same as saying that there is a SU(3,3) structure on the manifold. 
(If the spinor is also closed, Hitchin calls these
manifolds {\it generalized \cy s}.) 
For a SU(3) structure, there are {\sl two} pure spinors which are
orthogonal and of unit norm. From this point of view it is natural to
conjecture that mirror symmetry between two SU(3) structure manifolds 
exchanges these two pure spinors. We can be more explicit if we compare 
this \clss\ spinor definition of SU(3) structure with the more usual one, 
existence of a two--form $J$ and three--form 
$\Omega$ obeying $J\wedge \Omega=0$ and $i\Omega\wedge\bar\Omega = (2J)^3/3!$.
In these terms the two pure spinors are $e^{i J}$ and $\Omega$. We can
actually multiply first spinor by $e^B$ leaving it pure \cite{hitchin}.
So what we are claiming is 
\begin{equation}
  \label{eq:ejom}
e^{B+iJ}\longleftrightarrow \ \Omega  \ .
\end{equation}
The arrows here will be made precise in section \ref{sec:Mirror}.
In the \cy\ case, this exchange is implicit in many applications of
mirror symmetry. For example, the even periods 
and the D--brane charge can be written using $e^{B+i J}$, and its exchange
with $\Omega$ was used in mapping \cite{lyz} stringy--corrected DUY equations
\cite{mmms} to the special lagrangian 
condition; $e^{B+i J}$ was also used in formulating  the concept of
$\Pi$--stability \cite{dfr}.

With this in mind, we check that
T--duality along $T^3$ (when it is possible) realizes the exchange
(\ref{eq:ejom}), and for this reason we call it mirror symmetry. In this sense
we have generalized part {\it ii}) of SYZ.  However as it stands, (\ref{eq:ejom}) is hardly useful in predicting the mirror background starting from a particular six-manifold and NS flux.

After having discussed and justified the method, we can schematically describe
here our results. The usual 
quantities which measure non--Ricci--flatness of the
SU(3) manifold are the five components of the intrinsic torsion (mentioned above) labeled as
$W_i$, $i=1\ldots 5$, in the representations $1\oplus1$, $8\oplus8$,
$6\oplus\bar6$, $3\oplus\bar3$, $3\oplus\bar3$ respectively. What is puzzling
at first is that one does not see many ways of mirror pairing these
representations, except for $W_4$ and $W_5$ which are two vectors. The answer
is that the two mirrors have indeed two different SU(3) structures: the two
SU(3) are differently embedded into Spin(6,6), because the fiber directions
change from tangent bundle to cotangent bundle, roughly speaking.
As a result, representations get actually mixed. What is preserved is
the representations that these objects have once pulled back to the base
manifold, which is untouched by T--duality. $W_2$ and $W_3$ get then split
as $W_2=w_2^s+w_2^a$ ($8\to 5 \oplus 3$) and $W_3=w_3^s+w_3^t$ ($6\to
5 \oplus 1$),
and we get
\begin{equation}
  \label{eq:wtrans}
  \begin{array}{c}
  W_1 - iH_1  \longleftrightarrow -\overline{(W_1 - iH_1)}\ , \\[2mm]
  \bar{w}_2^s  \longleftrightarrow w_3^s - ih_3^s, \\[2mm]
  w_5, \bar{w}_2^a \longleftrightarrow w_4 - ih_4\ .
  \end{array}\end{equation}
A more detailed discussion of these equations can be found in section \ref{sec:W}
(see in particular (\ref{eq:tw}) for the precise statement).
In (\ref{eq:wtrans}) one can  see that $W_1$, $W_3$, $W_4$ get naturally
complexified by the components of $H$ in the corresponding
representation. This is no surprise as these torsions appear, as we review
later, in $dJ$, and the natural object in string theory is always $B+i J$. In
the present context, this arises rather due to the usual combination
$E=g + B$ of T--duality. As we will specify in section \ref{sec:geometry}, we
mostly work with a purely base--fiber type $B$--field, which is not the
most general form
allowed by $T^3$ invariance. However, we will see that this is just a simplifying
technical assumption, and may eventually be relaxed.
Note also that (\ref{eq:wtrans}) complements (\ref{eq:ejom}) in an essential
way by specifying in a more practical fashion the data of the mirror background (the metric and the NS flux). In particular, it quantifies the exchange of components of the flux and the intrinsic torsion on mirror sides.

We have also indicated in (\ref{eq:wtrans}) that some of the components of
the intrinsic torsion we begin with are actually related; so $T^3$
fibrations
are not the most general SU(3) manifold. This in a way answers in the negative
the question about generalizing part {\it i}) of SYZ: we are getting 
mirror symmetry only for a subclass of manifolds. In particular,
supersymmetric vacua with only the three-form switched on are outside 
this class: indeed the conditions
for these are \cite{strominger} (reinterpreted in terms of torsions for
example in \cite{britpop} and  \cite{berlin})
\begin{equation}
  \label{eq:hbg}
  W_1=W_2=0\ ,\quad W_3= *H_3\ ,\quad
  W_4=d\phi=iH_4\ , \quad W_5=2d\phi
\end{equation}
where we denoted by $H_4$ and $H_3$ the components in the
representations $3$ and $6$ of SU(3) respectively, in analogy with the
notation for torsions (see also the appendix \ref{app:torsion}). So the
next
natural step would be to try and include more general classes, among which
maybe supersymmetric vacua\footnote{There may actually be 
supersymmetric vacua involving $T^3$ fibrations, if other fields are 
included.}. In order to do so, it is natural to wonder to
what extent the transformation rules can be put in a nicer form, and in
particular be covariantized.
It turns out that it is convenient to use spinors. Although also the
previously mentioned \clss\ spinors can be used, here we mean a
more conventional Clifford(6) spinor without zeros. One such spinor, call it
$\epsilon$, always exists on any SU(3) structure
manifold and can actually be used to define it. It turns out that using
$\epsilon$ a different basis for $W$'s can be defined, which is {\sl
diagonal} 
under T--duality: elements of the basis transform picking a sign. 

The idea of 
this different spinorial   basis for $W$'s is roughly speaking the
following. Usual $W$'s are
defined, as we review later, from $dJ$ and $d\Omega$. Now, not only $\epsilon$
is equivalent to the pair $J, \Omega$, but the information contained
in $dJ$ and $d\Omega$ can also be completely
extracted from $D_M \epsilon$. Using SU(3) structure, this can be
decomposed as $D_M \epsilon =
\left(q_M + i \tilde q_M \gamma + i q_{MN} \gamma^N\right)\epsilon,$
where $\gamma$ is the chiral gamma in six dimensions and the group representations
inside the quantities $q_M$, $\tilde q_M$, $q_{MN}$ are in one-to-one
correspondence with the $W$'s.
Switching to the spinorial basis  accomplishes two things. First, it
allows to capture the exchange of the pure spinors $e^{iJ}$ and $\Omega$
and the exchange of their integrability properties simultaneously. More
importantly, it allows to conjecture the six-dimensional covariantization
of the mirror transformation (\ref{eq:wtrans}), written in terms of the
forms pulled back to the base of the $T^3$ fibration. Details can be found
in section \ref{sec:Q}.

For the purposes of studying mirror symmetry/T--duality we will need
first
to introduce the covariant derivative twisted by the NS flux:
\begin{equation}
 \label{eq:dirh}
  D^H_M \epsilon =
\left(Q_M + i \tilde Q_M \gamma + i Q_{MN} \gamma^N\right)\epsilon
\end{equation}
where, as we will see in detail, $Q$'s are
obtained from $q$'s by complexifying certain components of the intrinsic
torsion by the matching components of the flux (as in (\ref{eq:wtrans})).
We will show that their restrictions to the base (denoted by hatted
quantities) transform as
\begin{equation}
  \label{eq:qtrans}
  \hQ_{ij} \longrightarrow -\bar{\hQ}_{ij}, 
\qquad
  \hQ_i    \longrightarrow -\bar{\hQ}_i.
\end{equation}
We will then argue that
this simplification is due to the simple transformation of the
ten--dimensional spinors  under T--duality. 

Finally we will try, in section \ref{approaches}, to
collect these several points of view to argue that in general a rule like
$$
  Q_{mn} \longleftrightarrow - Q_{m\bar n}\ , \qquad
  Q_m \longleftrightarrow -{\bar Q}_m
$$
should hold. 
This rule is consistent with what we found in the $T^3$ fibered case, and
with the principle
that supersymmetric vacua should map in supersymmetric vacua (not necessarily
the same). There are however more checks that could be done if one
understood better examples; we discuss this in section \ref{ae}. For
example, in the case we mentioned above of compactifications with $H$
only described by (\ref{eq:hbg}), one should understand moduli spaces
and then check that a kind of exchange of complex and K{\"a}hler moduli
(although, as we will argue, this has to be taken with a grain of
salt) should happen. This might be interesting for the problem of
fixing moduli. We end with a section on open problems.

\section{Geometric setting} \label{sec:geometry}
We start with an introductory section on T--duality, mainly to fix
the notations.

The six--dimensional manifold will be taken to be a $T^3$ fibration over a
base $B$. Coordinates on the base will be denoted by $(y^1, y^2, y^3)$, and on
the fiber by $(x^1, x^2, x^3)$. 
All the quantities will only depend on the
$y$ coordinates, so that the $x$ directions are Killing
vectors.
We will use conventions for indices as follows:
\begin{itemize}
\item $i,j,k,\dots$ are used in the 3d $y$ subspace,
\item $\alpha,\beta,\gamma,\dots$ are used in the 3d $x$ subspace,
\item $M,N,\dots$ are used in the total 6d space for real coordinates~: $dy^M=(dy^i,dx^\alpha)$,
\item $m,n,\dots$ are used for holomorphic/antiholomorphic indices,
\item $A,B,C,\dots$ are indices in the total 3d complex frame space,
\item $a,b,c,\dots$ and $a',b',c',\dots$ are used in the 3d real $y$ and
  $x$ frame spaces. Primes will be dropped quickly.
\end{itemize}

We write then the most general metric and B field as \footnote{with the convention
  that $\omega_1 \wedge \omega_2 = \omega_1 \otimes \omega_2 -
  \omega_2 \otimes \omega_1$ }
\begin{eqnarray}
ds^2 &=& g_{ij}\, dy^i dy^j + h_{\alpha\beta}\,e^\alpha e^\beta
      =  G_{MN} dy^M dy^N \label{eq:metric}\\
 B_2 &=& \frac{1}{2} B_{ij}\, dy^i \wedge dy^j 
         + B_\alpha \wedge (dx^\alpha + \frac{1}{2}\, \lambda^\alpha)
         + \frac{1}{2} B_{\alpha\beta}\, e^\alpha \wedge e^\beta \label{eq:Bfield}
\end{eqnarray}
where $\lambda^\alpha = \lambda_i^\alpha dy^i$, 
$B_\alpha = B_{i\alpha} dy^i$ and we have defined 
$$
e^\alpha \equiv dx^\alpha + \lambda^\alpha\ .
$$
Of course the vielbein reads $(e^a_i dy^i, V^a_\alpha e^\alpha)$, where
$$ \begin{array}{rcl}
     \delta_{ab} e^a_i e^b_j &=& g_{ij}\ ,\\[2mm]
          g^{ij} e^a_i e^b_j &=& \delta^{ab}\ ,\\
   \end{array}
\qquad\qquad
   \begin{array}{rcl}
       \delta_{a'b'} V^{a'}_\alpha V^{b'}_\beta &=& h_{\alpha\beta}\ ,\\[2mm]
     h^{\alpha\beta} V^{a'}_\alpha V^{b'}_\beta &=& \delta^{a'b'}\ ;\\
   \end{array}
$$
we also record that the inverse vielbein has instead the form 
\begin{equation}
  \label{eq:invviel}
e_a^i(\frac{\de}{\de y^i} - \lambda_i^\alpha\frac{\de}{\de x^\alpha})\ ,\qquad 
V_a^\alpha\frac{\de}{\de x^\alpha}\ .  
\end{equation}

T--duality along the three $x^\alpha$ directions can be expressed conveniently 
in terms of the quantity $E= g +B$:
\begin{equation}
  \label{eq:tduality}
  \begin{array}{c}\vspace{.3cm}
  E_{ij}dy^i dy^j + E_{i\alpha}dy^i dx^\alpha +
E_{\alpha i}dx^\alpha dy^i + E_{\alpha\beta}dx^\alpha dx^\beta \\
\mapsto E_{ij}dy^i dy^j + E^{\alpha\beta}(dx^\alpha +E_{i\alpha} dy^i) 
(dx^\beta +E_{\beta j} dy^j) \ ;
  \end{array}
\end{equation}
notice that in this expression all the (implicit) tensor products are neither
symmetrized nor antisymmetrized, for example $dy^i dy^j = dy^i\otimes dy^j$.
Also remark that in this expression we used $dy^i, dx^\alpha$ basis 
instead of  $dy^i, e^\alpha$ as virtually everywhere else. 
$E^{\alpha \beta}$ is the inverse of $E_{\alpha\beta}$ and can be decomposed 
in symmetric and antisymmetric part: 
\begin{equation} \label{eqn:invhB}
  E^{\alpha\beta} = \left( \frac{1}{h+B} \right)^{\alpha\beta}
    = \hat{h}^{\alpha\beta} + \hat{B}^{\alpha\beta}
\qquad\mbox{where }
\left\{ \begin{array}{ccc}
\hat{h} &=& {\displaystyle \frac{1}{h+B}\, h \,\frac{1}{h-B}}\\[3mm]
\hat{B} &=& {\displaystyle \frac{1}{h+B}\, (-B) \,\frac{1}{h-B}}\\
\end{array} \right.
\end{equation}
The objects $\hat h$ and $\hat B$ would also called in other contexts
$H$ and $\Theta$. (In this paper $H$ denotes instead the three--form field.)

Using the relations 
$$ 
\hat{B} \hat{h}^{-1} = - h^{-1} B
\ ; \qquad
   \hat{h}^{-1} \hat{B} = - B h^{-1}
\ ; \qquad
   \hat{B} \hat{h}^{-1} \hat{B} = \hat{h} - h^{-1}
$$
one can show that the T--dual metric and $B$ field can be obtained by the
original ones (\ref{eq:metric}), (\ref{eq:Bfield}) by the substitutions
\begin{equation}
  h_{\alpha\beta} \longleftrightarrow \hat{h}^{\alpha\beta}
\  ; \qquad
  B_{\alpha\beta} \longleftrightarrow \hat{B}^{\alpha\beta}
\ ; \qquad
  B_\alpha \longleftrightarrow \lambda^\alpha 
\ .\label{eq:Td}
\end{equation}
and leaving the $g_{ij}$ and $B_{ij}$ in variant.
Notice that last equation in (\ref{eq:Td}) means that the twisting of each
of the three $S^1$ bundles gets exchanged with $B$ field. This fact played
for example a role in a number of applications and was recently formalized in mathematical 
terms in \cite{bem}.

We can also find the vielbein $\hat{V}^{a\alpha}$ of the T--dual metric
$\hat{h}^{\alpha\beta}$, that satisfies
$\hat{V}^{a\alpha} \hat{V}^{a\beta} = \hat{h}^{\alpha\beta}$:
\begin{equation}
  \hat{V}^{a\alpha}
 = \left( \frac{1}{h+B} \right)^{\alpha\beta} V^a_\beta
 = V^a_\beta \left( \frac{1}{h-B} \right)^{\beta\alpha}
\end{equation}
whose inverse is
\begin{equation}
  \hat{V}^a_\alpha
 \equiv \hat{h}^{\alpha\beta} \hat{V}^{a\beta}
 = (h-B)_{\alpha\beta} V^{a\beta}
 = V^{a\beta} (h+B)_{\beta\alpha}\ .
\end{equation}
The T--duality transformations of the vielbeine then are:
\begin{equation}
\label{eq:Tviel}
 V^a_\alpha  \longleftrightarrow \hat{V}^{a\alpha}
\  ; \qquad
 V^{a\alpha} \longleftrightarrow \hat{V}^a_\alpha .
\end{equation}

We will mostly work in the case when the $B$-field is purely of
base--fiber
type in frame indices. Transformation (\ref{eq:Td}) shows that this
condition is conserved by T--duality, while (\ref{eqn:invhB}) reduces to
$\hat{h}^{\alpha\beta}=h^{\alpha\beta}$. Consequently,
$\hat{V}^{a\alpha}=V^{a\alpha}$ and $\hat{V}^a_\alpha=V^a_\alpha$.
T--duality then only amounts to moving fiber indices up and down (still
exchanging $B_\alpha$ and $\lambda^\alpha$ though).

For later use, we also define here the tensors defining the SU(3) structure.
These would be a priori only a two--form $J$ and a three--form $\Omega$
satisfying $J\wedge\Omega=0$ and $i\Omega\wedge\bar\Omega =(2J)^3/3!$, but here we
define the structure in a more conventional way starting from an almost
complex structure. The latter is defined by giving the $(1,0)$ vielbein
\begin{equation}
  \label{eq:holviel}
E^A = i e^a_i dy^i + V^{a'}_\alpha e^\alpha
\end{equation}
where $A=a=a'$ goes from 1 to 3.
The corresponding (0,1) vielbein is $E^{\bar{B}} = \overline{E^B}$.
This almost complex structure is in general not integrable, as (even after
rescaling) it is not expressible as $d$ of a complex
coordinate, $E^A\neq \alpha^A dz^A$. However, with an abuse of language
we will use the quantity
\begin{equation}
  dz^j \equiv dy^j - i V^j_\gamma e^\gamma = -i e^j_a E^a,
\end{equation}
keeping in mind that there is no reason for an actual coordinate $z^i$ to exist.
We also used in this expression the notation
$$ V_{i\alpha} \equiv \delta_{aa'} e^a_i V^{a'}_\alpha
               = e^a_i V_{a\alpha} $$
%
%
The two--form $J$ (sometimes called {\em fundamental form}) is defined by
\begin{equation}
  \label{eq:J}
  J = \frac{i}{2}\: \delta_{AB}\: E^A \wedge E^{\bar{B}}
    = \frac{i}{2}\: g_{ij}\: dz^i \wedge d\bar{z}^j
    = -V_{i\alpha}\: dy^i \wedge e^\alpha
\end{equation}
%
The holomorphic 3-form reads instead
\begin{equation}
  \label{eq:omega}
  \Omega = E^1 \wedge E^2 \wedge E^3
         = \frac{1}{6}\: \epsilon_{ABC}\: E^A \wedge E^B \wedge E^C
         = -\frac{i}{6}\: \epsilon_{ijk}\: dz^i \wedge dz^j \wedge dz^k,
\end{equation}
where $\epsilon^{ijk} = \epsilon_{abc}\, e^{ai} e^{bj} e^{ck}$.

The choices we are making for the SU(3) structure are inspired by the
SYZ approach. As we stressed above, these choices reduce the structure
group further and are thus not to be expected to be as general (not
even locally) as the $T^3$ fibration structure was in the SYZ
approach. In particular some unaesthetic features will arise later
in the dual of the complex coordinates. Anyway, in section
\ref{approaches} we will try to amend to this loss of generality.


\section{Mirror symmetry as T--duality}
\label{sec:Mirror}

We start in this section showing, as promised in the introduction, how
$e^{B+iJ}$ and $\Omega$ get exchanged by T--duality.

First we do the easier case, in which there is no $B$ field and $\lambda$ 
twisting of the $T^3$ bundle. The basic idea is that $\Omega$ can be
written in a sense as an exponential of the almost complex structure 
${J_M}^N$ applied to a degenerate three--form $\epsilon_{ijk}dy^i dy^j
dy^k$, that can be thought of as the holomorphic three--form in the large
complex structure limit. A way to be more explicit is the following.  
Expand $\Omega$ from (\ref{eq:omega}) using the expression for the holomorphic
vielbein in (\ref{eq:holviel}). One obtains four terms, with $dy^3$, $dy^2
e$, and so on. Define now the operation $V^\perp\lrcorner(\cdot )$ by
\[
V^\perp \lrcorner(e^{\alpha_1}\ldots e^{\alpha_k}) = \frac1{(3-k)!}
\epsilon^{\alpha_1\ldots
  \alpha_3} e_{\alpha_{k+1}} \ldots e_{\alpha_3} \ , k=0\ldots 3\ .
\]
This is essentially a Hodge star on the fiber, except it sends a $k$-form
in the fiber into a $3-k$-vector (a section of $\Lambda^{3-k}T$).
Lower $e_\alpha$ are indeed vectors $\de_\alpha\equiv \de/\de x^\alpha$.
This operation is very similar to the T--duality 
transformation of spinors to be discussed shortly. Using this,
on every component of the expansion in $dy$ and $e$ of $\Omega$,
we get a sum of $(k,k)$ tensors, namely 
$k$ indices up and $k$ down: those down are along the fiber. The sum can
be
expressed as an exponent of $V_i^\alpha e_\alpha dy^i$, which is the complex
structure.  T--duality is now easy to perform.  According to 
(\ref{eq:Tviel}) its action is simply to raise and lower $\alpha$ index:
the tangent
bundle (in the fiber direction) of the starting manifold is equal to the
cotangent bundle (again in the fiber direction) of the T--dual manifold.
As a result the complex structure gets now mapped to $V_{i\alpha}e^\alpha
dy^i$, the fundamental two--form $J$.
So we have gotten 
\begin{equation}
  \label{eq:omejnob}
  T(V^\perp \lrcorner \Omega) = \frac i{3!}\: e^{i J}\ .
\end{equation}

The case with $B$--field and $\lambda$ is less trivial. Although this is not
strictly required here, we find already at this point helpful to think about
this in terms of \cldd\ spinors. So we make a brief intermezzo explaining 
these and then we get back to our computation. Much of this material is taken 
from \cite{hitchin}.

\subsection{\cldd\ spinors}

Clifford algebra is usually defined on the tangent bundle (or cotangent) of a
manifold using the metric. In physical notation this amounts to defining 
$d$ gamma matrices which satisfy $\{ \gamma^M, \gamma^N\}= 2 g^{MN}$, where
$g^{MN}$ is the metric on the cotangent bundle of the manifold. On SU(3)
manifold there is moreover a well--known representation of this Clifford
algebra, on $\Omega^{0,p}$ forms. 
If we on the contrary 
forget about the metric (thus about the SO(d) structure),
this algebra cannot be defined. 

If we consider, however, both the tangent and the cotangent bundles of the
manifold at the same time, there is a natural pairing between them (namely 
contraction between a vector and a form, $(dy^M , \de_N)={\delta^M}_N$), in
which the metric does not enter. This ``metric'' on $T\oplus T^*$ is
block--off--diagonal
\[
\left(\begin{array}{cc}
0&1 \\ 1&0
\end{array}\right)
\]
and thus of signature $(d,d)$. 
Concretely, what this means is that one has to define $2d$ {\sl independent }
gamma matrices,
$\gamma^M, \gamma_M$, that satisfy
$\{ \gamma^M, \gamma^N\}= 0=\{ \gamma_M, \gamma_N\}$ and
$\{ \gamma^M, \gamma_N\}= \delta^M_N$. Even though the Clifford structure has
been defined on $T\oplus T^*$, fortunately  the algebra still has a
representation in terms of the forms on the manifold. Only now we have
twice the number of creators and annihilators, and instead of using simply
$(0,p)$ forms as
before, we have to use forms of all possible degrees.
On this space $\oplus_{p=1}^d \Lambda^p T^*$, an explicit representation is
\begin{equation}
  \label{eq:repcldd}
  \gamma^M= dy^M\wedge \ , \qquad \gamma_M = \iota_{\de_M}\ .
\end{equation}
In all this we stress again that we have to consider $\gamma_M$ and $\gamma^M$
as independent: we cannot raise and lower indices using the metric.
In this \cldd\ algebra, however, the usual Clifford(d) is embedded: indeed 
a combination of wedge and contraction in (\ref{eq:repcldd}) is the more
conventional Clifford product, and if we use that we can raise and lower
indices. 

As stated in the introduction, a {\it pure} spinor is one which is annihilated
exactly by half of the gamma matrices. If we come back at the application we have
in mind, both $e^{i J}$ and $\bar\Omega$ are pure:
\begin{equation}
  \label{eq:pure}
  (\gamma^M -i J^{MN}\gamma_N)e^{i J}=0\ , \qquad
\begin{array}{c}\vspace{.2cm}
(\gamma^M- i {J^M}_N\gamma^N)\bar\Omega=0\ ,\\
(\gamma_M- i {J_M}^N\gamma_N)\bar\Omega=0\ .
\end{array}
\end{equation}
The gammas that annihilate the pure spinor $\bar\Omega$ 
are more familiar if one expresses them in
holomorphic/antiholomorphic indices: $\gamma^m \Omega=\gamma_{\bar m
  }\Omega=0$. Indeed $\bar\Omega$ is one of the Clifford vacua for the
Clifford(d) representation mentioned above (this is why we wrote the relations
for $\bar\Omega$ rather than $\Omega$). Let us also notice that the
annihilators of the two \cldd\ spinors in (\ref{eq:pure}) become the same
when we allow ourselves to raise and lower indices on gammas, that is, when 
we descend to Clifford(d): $e^{iJ}$ becomes then an alternative expression 
for a Clifford vacuum of Clifford(d).

As already mentioned, this dual way of realizing \cld\ from \cldd\ is
obviously in the center of mirror symmetry - exchange of the the K{\"a}hler
form and the holomorphic three-form (or their non-integrable
generalizations) is seen as different choices of Clifford vacuum.

\subsection{Back to mirror symmetry}
\label{back}

In this section, the only parts of the above theory that we actually use
here are the formulas for the annihilators (\ref{eq:pure}), which of
course could have been derived independently. This insight gives however
a useful rule
of thumb, in particular when
dealing with $e^{iJ}$, where we can save ourselves expanding the
exponential as we
did above. What we will do in the
following will be to consider $e^{iJ}$ and $\Omega$ as \clss\ spinors, and
other forms acting on them as combinations of gamma matrices. This is of course
not the only possibility. One might have included $B$ in the definition of
the pure spinor. Due to technical details in how T--duality works we preferred
this way. Also, we will work here in the case $B_{\alpha\beta}=0$.

Let us consider for example the expression $e^B \Omega$. Due to
$\gamma^\alpha\Omega = i \gamma^i V_i^\alpha\Omega$, this equals
$e^{iB_\alpha\wedge V^\alpha}\Omega$. If we act on this with the operator
$V^\perp\lrcorner$ defined above, the prefactor can be taken out (it does not
contain any $e^\alpha$). On $\Omega$ we get $V^\perp\lrcorner (\Omega) =
e^{-i V^\alpha e_\alpha } $ as above; the only thing to notice is that
$e_\alpha$ is simply $\de_\alpha$, as seen on (\ref{eq:invviel}).
If we finally apply T--duality, $V^\alpha e_\alpha \mapsto V_\alpha e^\alpha$;
putting it together with the inert factor $e^{iB_\alpha\wedge V^\alpha}=
e^{iB^\alpha\wedge V_\alpha}$, we have shown
\begin{equation}
  \label{eq:tdual1}
 \frac i{3!}  T(e^{i J}) = V^\perp \lrcorner (e^B\Omega)\ e^{-B_{\alpha}
\lambda^{\alpha}} .
\end{equation}
It is a little surprising that the $B$ field has to be subtracted on right
hand side rather than being already present on left hand side.
In the same way
we can also prove the more reassuring 
\begin{equation}
  \label{eq:tdual2}
  T(\Omega)= \frac i{3!} V^\perp \lrcorner (e^B e^{i J})\ e^{B_{\alpha}
\lambda^{\alpha}} .
\end{equation}

The exchange $e^{B+iJ}\longleftrightarrow \ \Omega$ as presented in
(\ref{eq:tdual1}) and (\ref{eq:tdual2}) is not very aesthetically pleasing,
however the exponents involving the T--duality anti-invariant $B_{\alpha}
\lambda^{\alpha}$ are easy to explain going back to (\ref{eq:pure}). The
condition of purity e.g. on
$\Omega$ is essentially $dz \wedge \Omega=0,$ and the holomorphic
coordinates  change under T--duality.
The reason of this is that the $dz^i$ which we have
defined above as $dy^i - i V^i_\gamma e^\gamma$ has a $\lambda$ hidden
inside $e^\gamma$. Since $\lambda$ gets exchanged with $B$ due to
(\ref{eq:Td}), $dz$ on the original manifold does not map exactly to
$dz$, but $dz
\longrightarrow dz -i(B_{\alpha} V^{\alpha} -
\lambda^{\alpha}V_{\alpha})$
shifting by another T--duality anti-invariant. Thus the role of $e^{\pm
B_{\alpha} \lambda^{\alpha}}$ is to compensate for this change, preserving
the condition for purity.

The combinations $e^{i J}$ and $\Omega$ allow, as we have commented on in the
introduction and as we will see further later
on, to treat $J$ and $\Omega$ more symmetrically. The most symmetrical object
one might imagine is actually the SU(3) invariant spinor $\epsilon$ itself.
Given also the role that we anticipated it will have in torsions, one 
might wonder at this point if it is more convenient to use T--duality
transformation of $\epsilon$ and forget all the rest. The problem is, so to
say, that the spinor is too symmetric. The transformation rule of the
ten--dimensional spinors are known: in the case without $B_{\alpha\beta}$,
we simply have $\psi_+ \to \psi_+$, $\psi_-\to \gamma_f \psi_-$, where
$\gamma_f$ is the product of the three gammas in the fiber directions
\cite{hassan}. 
However, when we express $\psi_\pm$ in terms of the chirality projected 
$\epsilon_\pm$ of the six--dimensional spinor, $\gamma_f \epsilon_+$ is 
actually $\epsilon_-$ and all the information we get
is that a IIA compactification has been exchanged with a IIB one. This means
that the spinor is essentially on both the original and the T--dual manifold
the pull--back of a spinor in the base. Still, using the familiar
bilinear definitions for $J$ and $\Omega$ (\ref{eq:bil}) and $\gamma_f
\epsilon_+=\epsilon_-$, one can show the identities above in a different way.

\section{Intrinsic torsions and their duals}
\label{sec:W}

This section is the technical core of the paper. Here we define and
compute intrinsic  torsions for our $T^3$ fibered manifolds. As stressed in the
introduction, these are not the most general SU(3) structure manifolds. 
Performing T--duality along the $T^3$ is then easy using (\ref{eq:Td}) and
(\ref{eq:Tviel}).

\subsection{Conventional definition of torsions}

We do not aim here at reviewing intrinsic torsions on manifolds with
G--structures as discussions already exist in the literature, see
for example \cite{joyce} and among recent physics papers 
\cite{glmw,britpop}. Here we give a good working definition. It is familiar
that, if we are on a SU(3) {\it holonomy} manifold, not only $J$ and $\Omega$
are well defined, but also they are closed: $dJ=0=d\Omega$. If they are not,
$dJ$ and $d\Omega$ give a good measure of how far the manifold is from having
SU(3) holonomy. The usual definitions require to split them in
SU(3) representations:
\begin{equation}
  \label{eq:djdo}
  \begin{array}{c}\vspace{.3cm}
dJ = -\frac{3}{2}\, {\rm Im}(W_1 \bar{\Omega}) + W_4 \wedge J + W_3\\
d\Omega = W_1 J^2 + W_2 \wedge J + \bar{W_5} \wedge \Omega\
  \end{array}
\end{equation}
where the representations of the $W_i$ are as follows:
\begin{itemize}
\item $W_1$ is a complex zero--form in $1 \oplus 1$;
\item $W_2$ is a complex primitive two--form, so it lies in $8 \oplus 8$;
\item $W_3$ is a real primitive $(2,1) \oplus (1,2)$ form, so it lies in
 $6 \oplus \bar{6}$;
\item $W_4$ is a real one--form in $3 \oplus \bar{3}$;
\item $W_5$ is a complex $(1,0)$--form (notice that in (\ref{eq:djdo}) the
  $(0,1)$ part drops out), so its degrees of freedom are again 
$3 \oplus \bar{3}$. 
\end{itemize}

These $W_i$ allow to classify quickly any SU(3) manifold. We will later define
them in an alternative way using directly the spinor; that definition will be
more natural for T--duality, but the $W$'s are often better to analyze the
type
of the manifold. For example, 
notice that in (\ref{eq:djdo}) the exterior derivative $d$ does not satisfy
the usual rule $d: \Omega^{p,q}\to \Omega^{p+1,q}\oplus
\Omega^{p,q+1}$. For an almost complex manifold 
as we have here, there are also $(p+2,q-1)$ and $(p-1,q+2)$
contributions.
Hence in (\ref{eq:djdo}) the $(3,0) \oplus (0,3)$ part of $dJ$, namely
Im($W_1 \bar{\Omega}$), and the $(2,2)$ part of $d\Omega$, which reads 
$W_1 J^2 + W_2 \wedge J$.
So we know actually that $W_1=W_2=0$ iff the manifold is complex. One can
check indeed that the Nijenhuis tensor can be expressed in terms of $W_1$ and
$W_2$. Other examples of the use of these $W$'s abound in the literature.
Notice also that the information of $dJ$ and $d\Omega$ is a little redundant,
as $W_1$ appears in both. 

Before we start computing, notice that from this classical definition it 
would be not obvious to guess transformation laws for $W$'s, other than
some
qualitative features. There are two vectors, but the $8$ and the $6$ are
different representations. If one thinks already at this stage about
decomposing in base representations, guessing becomes easier, but one feels
rapidly the need for a more solid ground. One way, which we pursue in this
section, is to compute blindly. The other way is to put $J$ and $\Omega$ on a
more symmetrical basis, using the formalism of \cldd, or, which is another
manifestation of the same idea, to actually use the SU(3) invariant spinor
directly. We do this in next section.

\subsection{Computations of torsions in the $T^3$ fibered case}

We can now compute $W$'s from the expressions (\ref{eq:J}) and
(\ref{eq:omega}). This is done by doing contractions, partial or total,
appropriate to isolate the component of interest. For example $W_4$ is
computed contracting $J\lrcorner dJ$.\footnote{Complete expressions   
for all five components of the intrinsic torsion for a metric of the form
(\ref{eq:metric}) can be found in appendix \ref{app:torsion}.}
First we give $W_1,W_4,W_5$, expressed in the holomorphic basis.\footnote{In what follows, we will denote $W$'s in complex coordinates as lower case $w$'s. For example, $\bar{w}^2_{ij}$ is $\bar{W}^2_{m\bar n}$, even if we did not explicitly mark $i$ and $j$ as holomorphic and antiholomorphic indices in this expression. This is also true for the other components.}
Note that $W_4$ is real and $W_5$ holomorphic, so that $W_4=w^4_i dz^i + \mathrm{c.~c.}$ and $W_5=w^5_i dz^i$, while $W_1=w_1$ is a scalar.\footnote{As already emphasized, one has to bear in mind that the almost complex structure is in general not integrable, so that $dz^i$ is not to be understood as the differential of a hypothetical coordinate $z^i$.}
These components read:
\begin{eqnarray}
  \label{eq:w1}
  w_1 &=& -\frac{i}{12}\, \epsilon^{ijk}\, V_{i\alpha}\;
           [ d(V-i\lambda) ]^\alpha_{jk} \\
\0 \\
  \label{eq:w4}
  w^4_k &=& -\frac{1}{4}\, V^{\alpha j}\, [d V_{\alpha}]_{jk} \\
\0 \\
  \label{eq:w5}
  w^5_k &=& -\frac{1}{4} \left\{ V^j_\alpha\, [ d(V+i\lambda) ]^\alpha_{jk}
                               - h^{\alpha\beta} \partial_k h_{\alpha\beta} \right\}
\end{eqnarray}
where $[d(\cdot)]_{ij}= 2\de_{[i}(\cdot)_{j]}$.

We now pass to $W_2$ and $W_3$. $W_2$ is a (1,1)-form, and $W_3$ is a real $(2,1) \oplus (1,2)$, and are written as
\begin{equation}
  \label{eq:w2w3}
  W_2 = w^2_{ij}\, dz^i \wedge d\bar{z}^j\ ,\qquad
  W_3 = \frac{1}{2}\, w^3_{ijk}\, dz^i \wedge d\bar{z}^j \wedge d\bar{z}^k + \mathrm{c.~c.} ,
\end{equation}
However, since the representation $6$ can be expressed not only as a primitive $(2,1)$ form, but also as a symmetric tensor with two holomorphic indices, we will give this latter expression for $W_3$. The way
to pass from one to another is $w^3_{ij}=w^3_{ipq}\Omega{^{pq}}_j$. This is
already a little in the spirit of the different basis for intrinsic torsion
that we will give later. Furthermore, these two matrices with indices $ij$ can actually be further decomposed in representation theory of the $SO(3)$ of the base. $w^2_{ij}$ has a symmetric and an antisymmetric part;
the symmetric part does not drop out, it only contributes to $dy\wedge e$
part; the antisymmetric part can be dualized to a three dimensional vector
$w_2^i = \frac{1}{2}\epsilon^{ijk}\, w^2_{jk}$.
As for $W_3$, $w^3_{ij} = w^3_{\{ij\}^0} + \frac{1}{3}\,w^3_t g_{ij}$ is
already symmetric but has a trace part $w^3_t$  on the three-dimensional
base (of course, it is  traceless in six dimensions).
\begin{eqnarray}
  \label{eq:w2}
  w^2_{\{ij\}} &=& \frac{1}{24}\,\epsilon^{pqk}\, [d(V-i\lambda)]^\alpha_{pq}
         [ 2 V_{k\alpha} g_{ij} - 3 V_{j\alpha} g_{ik} - 3 V_{i\alpha} g_{jk} ] \\
  w^2_k &=& -\frac{1}{4}  V^j_\alpha [ d(V-i\lambda) ]^\alpha_{jk}
\end{eqnarray}

\begin{eqnarray}
  \label{eq:w3}
  w^3_{\{ij\}^0} &=& \frac{1}{24}\,\epsilon^{pqk}\, [dV_\alpha]_{pq}
         [ 2 V_k^\alpha g_{ij} - 3 V_j^\alpha g_{ik} - 3 V_i^\alpha g_{jk} ] \\
\label{eq:w3t}  
w^3_t &=& \frac{1}{8}\,\epsilon^{pqk}\, [d(V-3i\lambda)]^\alpha_{pq}
V_{k\alpha}
\end{eqnarray}

Before turning to the T--duality transformations of components of the
intrinsic torsion and the flux, we observe that the conditions
for a supersymmetric vacuum with $H$ only (\ref{eq:hbg})
are not compatible in a nontrivial way with the expressions above, as we
anticipated in the introduction. For example, demanding $W_1=W_2=0$ sets
$\lambda$ and $V$ to constants.

\subsection{T--duality}

It is now easy to see what the transformation rules of the
$W$'s are. Decomposing in base representations says essentially where to
look.
One sees immediately that the various three--dimensional vectors and symmetric
matrices are all similar. Before spelling this out, one should however stress
that the full six--dimensional quantities have a more complicated
transformation rule. As explained in section \ref{back}, due to presence of
$\lambda$ in $e^\gamma$, $dz^i = dy^i - i V^i_\gamma e^\gamma$ on the
original manifold does not map exactly to $dz$ on the mirror side.

With this important caveat in mind, let us proceed to give T--duality
transformations. As we said, many of the expressions we have for $W$'s 
are similar (see appendix \ref{app:torsion}). The differences are mainly
because of $V_\alpha$ versus
$V^\alpha$. This is already good, as these quantities are exchanged by 
T--duality (\ref{eq:Tviel}). One also sees that some of the quantities contain
$\lambda$, that after T--duality become $B$ as we just recalled. So, we
are led naturally to complexify some of the torsions adding $dB$ projected
in the appropriate representation. As this projections are verbatim those 
we did for $dJ$ in previous subsection, this step is trivial.
Thus  defining components for $H$ as for other forms \footnote{The explicit
expressions for components of $H$ in the $T^3$-fibered geometry can be found in
appendix \ref{app:torsion}. We labeled these components so that they match the
corresponding ones in $dJ$. $H_1$ is then the $1 \oplus 1$ complex scalar, $H_3$
the $6 \oplus \bar{6}$ real 3-form and $H_4$ the $3 \oplus \bar{3}$ real 1-form.}
\begin{equation}
  \label{eq:compH}
  H = -\frac{3}{2}\, {\rm Im}(H_1 \bar{\Omega}) + H_4 \wedge J + H_3
\end{equation}
we find the transformations:
\begin{equation}
  \label{eq:tw}
  \begin{array}{c}
  w_1 - ih_1  \longleftrightarrow -\overline{(w_1 - ih_1)}\ , \\[2mm]
  w^2_{\{ij\}}  \longleftrightarrow (w_3 + ih_3)_{\{ij\}^0}, \\[2mm]
  w^5_k -\frac{1}{4}\, h^{\alpha\beta}\de_k h_{\alpha\beta} = \bar{w}^2_k
     \longleftrightarrow (w_4 - ih_4)_k\ .
  \end{array}
\end{equation}
describing the mixing of the components of the flux and of the intrinsic
torsion under mirror symmetry.

The central role in the mirror/T--duality transformation (\ref{eq:tw}) is
obviously played by $W_2$ (a component of the torsion associated with the
non-integrability of the complex structure).  It splits in two different
pieces upon restriction to the base and the respective mixing of the two
parts of $W_2$ with complexified $H_3$ and $H_4$ is an essential
ingredient of the mirror map.

We will now try  to rederive and generalize to generic geometries these
results from a different point of view, using spinors rather than
differential forms.

\section{Spinorial basis}
\label{sec:Q}

The idea is that the same information we have in $dJ$ and $d\Omega$ are
contained in $D_M \epsilon$. Doing the effort of reexpressing torsions in
these terms pays off for several reasons. First of all, the combinations that
appear in $D_M \epsilon$ transform better. Second, they might be useful in 
future occasions to analyze the geometry behind a given supersymmetry
transformation without even having to bother to construct bilinears. In
particular, we can find from this approach immediately the conditions 
(\ref{eq:hbg}) for supersymmetric vacua with $H$.

One proceeds in the following way. What we call $\epsilon$ in what follows is
the SU(3) invariant spinor, which can be furthermore decomposed by chirality 
as $\epsilon_+ +\epsilon_-$. 
Again, if we were on a manifold of SU(3)
{\it holonomy}, we would have a covariantly constant spinor, $D_M \epsilon=0$.
This is not the case, but still decomposing $D_M \epsilon$ into
representations will give us a measure of how far we are from SU(3) holonomy. 
The way of decomposing $D_M \epsilon$ into representations is again implicit
in the literature. On a SU(3) invariant manifold, a basis for spinors is
given by $\epsilon_\pm$ and $\gamma_M \epsilon_\pm$ (or alternatively we can 
trade $\epsilon_\pm$ with $\epsilon$ and $\gamma\epsilon$). So, for example, 
anything else in Clifford algebra acting on $\epsilon$, say $\gamma^{M_1\ldots
  M_n}$, can be reexpressed in terms of this basis. Explicit formulas for this
are known (see for example \cite{mmms}; in \cite{kmt} a complete set of these
equations are provided, along with the simple group theoretical description of
how to get them, for the case of seven--manifolds with G$_2$ structure). We
will not however need them here, it is enough to know that this
decomposition can be done. Actually, with one exception: the relation
$\gamma^M \gamma \epsilon = i {J^M}_N \gamma^N\epsilon$ can be used to
eliminate one possible term.  So we can write in general 
\begin{equation}
  \label{eq:deps}
  D_M \epsilon = \left(q_M + i\tilde q_M \gamma + iq_{MN}\gamma^N\right) \epsilon\ .
\end{equation}
The real $q$'s that we have {\sl defined} in this equation are just
another definition of intrinsic torsion. To see that they can be compared
with 
the $W$'s above, it suffices to use group theory. $q_M$ and $\tilde q_M$
are
vectors, $3 \oplus \bar 3$ ; as to $q_{MN}$, it can be decomposed into 
$(3 \oplus \bar 3)^{\otimes 2}= (6 \oplus\bar 3)\oplus(\bar 6 \oplus 3)
                         \oplus (8 \oplus 1) \oplus (8 \oplus 1)$.
We see that all the representations of the $W$'s are present. There is one
redundancy, since we get three vectors ($q_M$, $\tilde q_M$ and one from
$q_{MN}$). The objects we get in this way are the same as the $W$'s up to
factors. Qualitatively we could stop here; in the present context we are
actually interested in getting the factors, as they are important for
being able to express $q$'s in terms of  $W$'s explicitly. This is
done as follows. After having
decomposed $q_{MN}$ as above, we can define $J$ and $\Omega$ as bilinears
as
\begin{equation}
  \label{eq:bil}
  \epsilon^\dagger \gamma_{MN}\gamma\,\epsilon = i J_{MN}\ , \qquad
-i \epsilon^\dagger \gamma_{MNP}(1+\gamma)\epsilon = \Omega_{MNP}\ .
\end{equation}
One can now compute their exterior derivative using (\ref{eq:deps}).
Comparing the result with (\ref{eq:djdo}) gives the desired coefficients.
The result is 
\begin{eqnarray}
q_{MN} &=& \frac14\left(W_1^+ G_{MN} +W_1^- J_{MN} \right)
  +\frac18\left(\Omega_{MNP}\overline{(W_5-2P W_4)}^{P}
  + \mathrm{c.~c.}\right) \0 \\
       & & \phantom{xxx}
  + \frac14 \left( -{J_M}^P W^{2,+}_{PN} + W^{2,-}_{MN}\right)
  + \frac18\, {\rm Im}(W^3_{MN}) \0 \\
       &=& {\rm Re}\left[
   \frac{1}{2}\, W_1 \bar{P}_{MN}
  +\frac{1}{4}\, \Omega_{MNP} (\bar{W}_5 - 2W_4)^P
  +\frac{i}{2}\, \bar{P}{_M}^P W^2_{PN}
  -\frac{i}{8}\, W^3_{MN} \right] \\
q_M + i\tilde q_M &=& [\bar{W}_5 - \bar{P}W_4]_M
\label{eq:qmn}
\end{eqnarray}
where $W_i= W_i^+ -i W_i^-$ as usual in the literature, and we have
defined $W^3_{MN} = W^3_{MPQ} {\Omega^{PQ}}_{N}$ and used
a holomorphic projector $P=\frac{1}{2}(1-iJ)$. We had observed in the previous section
that the split of $W_2$ in two parts upon restriction to the base is
crucial in the mirror transformation. Here we can see that the split
nature of $W_2$ reveals itself in covariant six-dimensional expressions:
$W_2^{+}$ and $W_2^-$ enter respectively into the symmetric and
antisymmetric parts of $q_{MN}$.

It is worth recording the same expression in
holomorphic/antiholomorphic basis:
\begin{equation}
  \label{eq:qmnhol}
  q_{mn} = -\frac{i}{16} w^3_{mn} + \frac18\Omega_{mnp}(w_5 - 2w_4)^p
  \ , \qquad
  q_{m \bar n} = -\frac i4 \bar{w}^2_{m\bar n} +\frac14 \bar w_1 g_{m\bar n}\ .
\end{equation}
And for the remaining vector: 
\begin{equation}
  \label{eq:vec}
  q_m - i \tilde q_m = (w_5 - w_4)_m\ .
\end{equation}

The quantities we have defined so far would not be expected to behave
nicely under T--duality, for the following simple reason. The
transformation
laws we have computed in (\ref{eq:tw}) have, as one would expect also
from the arguments in \cite{glmw} and from (\ref{eq:ejom}), the
feature of exchanging some torsions with $H$. Therefore we have to add
a dependence on $H$ to the covariant derivative in (\ref{eq:deps}).
Then also the $q$ defined in (\ref{eq:deps}) will change and
(\ref{eq:deps}) will become
\begin{equation}
  \label{eq:dheps}
  D_M^H \epsilon = \left(Q_M + i\tilde Q_M \gamma +
    iQ_{MN}\gamma^N\right) 
\epsilon\ .
\end{equation}
We have defined $D^H$ (and as a consequence the $Q$'s) in such a way
as to find good T--duality transformation properties afterwards. Not
too  surprisingly, we have found that the best definition is exactly
the same as the one which appears in supergravity supersymmetry
transformations: $D_M^H\equiv
(D_M+\frac{1}{8} H_{MNP}\gamma^{NP})$. We find then
\begin{eqnarray}
  \label{eq:QMN}
  Q_{MN} &=& {\rm Re}\left[
   \frac{1}{2}\, (W_1 +3i H_1) \bar{P}_{MN}
  +\frac{1}{4}\, \Omega_{MNP} (\bar{W}_5 - 2(W_4 +i H_4))^P
   \right. \\ && \phantom{xxx} \left.
  +\frac{i}{2}\, \bar{P}_M^{\ P} W^2_{PN}
  -\frac{i}{8} (W^3 +iH^3)_{MN} \right]  \ .\0
\end{eqnarray}
So, adding $H$ as $D\to D^H$ complexifies $W$ as  $W+iH$, though at
the end the Re in (\ref{eq:QMN}) makes the $Q$'s real.
It should also be possible to write directly a formula for the
(con)torsion, as an alternative to formulas for the $q$'s that we
have
given. The fact that $H$ appears as $H_{MNP}\gamma^{NP}$ tells us
already that this formula will have a piece $K_{MNP}=dJ_{MNP}+\ldots$
that will combine with $H$. As we will not need it here, we do not
pursue this. Notice also that the G$_2$ analogue of what we just did
for $H$ is discussed in detail in
\cite{kmt} for the $G_2$ case.

The fact that the natural combination
for T--duality and for supersymmetry is the same will be useful later,
when we will try to extend our results to the general case. Then this is
also a good place to see that of course the conditions for
supersymmetry in the case with $H$ only (\ref{eq:hbg}) 
can be recovered from the spinor equation.
To have supersymmetry it is enough that one chirality, say $\epsilon_+$, is 
annihilated by $D^H$. We have the expressions
\begin{equation}
  \label{eq:susy}
  \begin{array}{rclcrcl}\vspace{.3cm}
  D^H_m \epsilon_+ &=& (Q_m + i \tilde Q_m) \epsilon_+
                     + i Q_{mn}\gamma^n\epsilon_- &\qquad&
  D^H_m \epsilon_- &=& (Q_m - i \tilde Q_m) \epsilon_-
                     - i Q_{m\bar n}\gamma^{\bar n}\epsilon_+ \\
  D^H_{\bar m} \epsilon_+ &=& (Q_{\bar m} + i \tilde Q_{\bar m}) \epsilon_+
                            + i Q_{\bar m n}\gamma^n \epsilon_- &\qquad&
  D^H_{\bar m} \epsilon_- &=& (Q_{\bar m} - i \tilde Q_{\bar m}) \epsilon_-
                            - i Q_{\bar m \bar n}\gamma^{\bar n} \epsilon_-
  \end{array}
\end{equation}
Notice that $Q_{m\bar n}$ and $Q_{\bar m \bar n}$ have disappeared
from $D^H \epsilon_+$,
because $\epsilon_-$, being a Clifford vacuum, is annihilated by
$\gamma^{\bar n}$. 
From this one obtains directly that the complexified $Q_{mn}$ and 
$Q_{\bar m n}$ have to vanish. These will say that the complexified
$W_3$ has to be purely antiholomorphic, which in more usual terms means 
of type $(1,2)$ (this is the condition $W_3= *H_3$) and that $W_2$ has to
vanish. The vectors require a little more care because usually the dilaton 
is rescaled in the metric (as a warping) and in the spinor itself. 
More generally it is clear that one can use gamma matrices identities
mentioned above to reduce the expression to a form like (\ref{eq:deps}), and
then use (\ref{eq:qmn}) or (\ref{eq:qmnhol}).

For us the main advantage of having computed these quantities is to compare
with T--duality transformations given in previous sections, although
we will see shortly how these supersymmetry considerations can play a role
in understanding the general case (without $T^3$ fibration structure).
We can restrict the free index in (\ref{eq:deps}) to be on the base, $M=i$, and furthermore
apply a chirality projector
\begin{equation}
  \label{eq:baseeps}
  D_i \epsilon_+ = \hQ_i \epsilon_+ + i \hQ_{ij}\gamma^j\epsilon_-\ ,\qquad
  D_i \epsilon_- = \bar{\hQ}_i \epsilon_- + i \bar{\hQ}_{ij}\gamma^j\epsilon_+\ .
\end{equation}
having introduced hatted quantities for restrictions to the base.
The quantities $\hQ_i$ and $\hQ_{ij}$ in these expressions turn out
to transform neatly under T--duality:
\begin{equation} \label{tdual_hQ}
  \hQ_i    \longrightarrow -\bar{\hQ}_i, \qquad
  \hQ_{ij} \longrightarrow -\bar{\hQ}_{ij},
\end{equation}
with the expressions
\begin{eqnarray}
  \label{eq:tinvs}
  \hQ_i &=& Q_i + i \tilde{Q}_i = (\bar{W}_5 - \frac{1}{2}\, (W_4 -iH_4))_i \\
  \hQ_{ij} &=& Q_{ij} - iQ_{i\alpha}V^\alpha_j = 2 \bar{P}_j^M Q_{iM} \0 \\
        &=& \frac{1}{4} \left[ \bar{W}_1 + 3i\bar{H}_{(1)}
                + \frac{i}{12}\, \bar{P}^{iM} (W_3 - iH_3)_{Mi} \right] g_{ij}
          - \frac{i}{4} \left[ \bar{W}^2_{\{ij\}}
                - \frac{1}{4} {\bar{P}_i}^M (W_3 - iH_3)_{Mj} \right] \0 \\
        & & \phantom{xxx}
          + \frac{i}{2}\, \epsilon_{ijk}
            \left[ W_5 - W_4 - iH_4 - \frac{1}{2}\, \bar{W}_2 \right]^k
\end{eqnarray}
This means that at an effective level the rule tells us
$\epsilon_+\leftrightarrow \epsilon_-$. 

We should remark that working so far with a finite-size $T^3$-fiber,
we have extra (nowhere vanishing) vector fields, and thus reduces
structure. This may in particular allow to locally preserve
supersymmetry even when conditions  (\ref{eq:hbg}) are violated.
Since when fibers degenerate this restricted stricture no longer
exists, we avoided making explicit use of it, even though doing
restrictions to the base manifolds implicitly uses the existence of a
restricted structure. It is reasonable to expect that the results
based on representations are valid over the entire moduli space, and
thus next we turn to the six-dimensional covariantization of mirror
transformation (\ref{eq:tw}).

\subsection{Approaches to the general case}
\label{approaches}

At this point it is natural to wonder if we have enough information 
to simply guess what mirror symmetry should be in the general case. 
We have a precise set of transformation rules in the case of $T^3$ fibrations,
and we also know that supersymmetric vacua should be sent to
supersymmetric vacua. As we remarked above, T--duality is induced by an
exchange of $\epsilon_+$ with 
$\epsilon_-$. Since we also have $\gamma^{\bar m}\epsilon_-=0$, 
these two facts together would suggest following proposal naturally
generalizing  
(\ref{tdual_hQ}):
\begin{equation}
  \label{eq:prop}
  Q_{mn} \longleftrightarrow - Q_{m\bar n}\ , \qquad Q_m \longleftrightarrow
  -{\bar Q}_m\ .
\end{equation}
We noticed above that representations of $W$'s do not match 
in such a way as to suggest immediately a transformation law. In the
T--duality approach above this was solved by decomposing further in
representations of the SO(3) of the base. The proposal
(\ref{eq:prop}), on the contrary, 
gets around this problem collecting together SU(3) representations
rather than decomposing them further: qualitatively, $6 \oplus \bar 3
\leftrightarrow 8 \oplus 1$.

Let us now check that this proposal for mirror symmetry agrees with T--duality
and with supersymmetry, as we just required. First of all,
(\ref{eq:prop}) agrees with the exchange (\ref{tdual_hQ}). Indeed
we have
\begin{equation}
 \hQ_{M\bar n} = {P_m}^P \hQ_{P\bar n} + {\bar{P}_{\bar m}}^P \hQ_{P\bar n}
               = 2 Q_{m\bar n} + 2 Q_{\bar m\bar n}\ ;
\end{equation}
similarly one can consider the transformation of $Q_m=\hQ_m$. 

Turning now to supersymmetry, the two transformations in
(\ref{eq:prop}) induce simply
\begin{equation}
  \label{eq:dhm}
D^H_m \epsilon_+ \longrightarrow -D^H_m \epsilon_- \ .
\end{equation}
So if only $H$ is present we are
sending $D^H \epsilon_+=0$ to $-D^H \epsilon_-=0$; in the latter case
supersymmetry is of course still preserved. In this form the duality
might seem a little tautological, in the sense that it sends a
supersymmetric vacuum in another one in an obvious way. Compare
however with the usual mirror symmetry: a \cy\ is sent to another \cy,
and the nontriviality lies in the exchange of K{\"a}hler and complex
structure moduli. This should be happening for vacua with $H$ only as
well, and in a sense this would be yet another check to do; we will
comment on this in next section.

Coming back to checking compatibility with supersymmetry, 
the situation becomes
more complicated with RR fluxes, because the latter also
transform, and one would have to check that they do it in a way
compatible with the one we are giving for geometry and $H$. 
This can be elaborated as follows. Just as the entire NS contribution to
the covariant derivative of the invariant spinor got summarized in $Q$'s
(see (\ref{eq:QMN})), the RR contribution can be accounted by introduction
of similar objects, $R_M$, ${\tilde R}_M$ and $R_{MN}$ with a group
decomposition matching that of $Q$'s. On supersymmetric backgrounds, the
total action of the covariant derivative of the invariant spinor should be
zero and thus $R=-Q$. Thus from this point of view the mirror
transformation of the RR sector can 
also be brought to the form (\ref{eq:prop}). From other side, 
in the $T^3$ fibered case, one could use the known transformation rules
of RR fields. 
From the above, it is clear that the natural way to do this check in
general would be to consider RR fields not as sums of forms but as
bispinors, expressing for example in terms of the latter also
supersymmetry transformations.

Even after all these motivations, the proposal (\ref{eq:prop}) stands
as a conjecture, and there would be other possible checks to be made.
One possibility is to use again the formalism of Clifford(6,6)
spinors.  One can give 
an alternative definition of torsions, that we have not mentioned
so far, using the \clss\ spinors $e^{i J}$ and $\Omega$. Schematically
one gets
\begin{equation}
  \label{eq:torscldd}
  D_M e^{i J} = q_M e^{i J} + Im( q_M^{(2)}\cdot \Omega)\ ,\qquad
  D_M \Omega = (q_M +i \tilde q_M)\Omega + q_M^{(2)}\cdot e^{i J} \ .
\end{equation}
In these equations, $q_M^{(2)}\cdot$ is the Clifford product of $q_{MN}$ using
only second index. These formulas seem indeed to be consistent with the
general rule (\ref{eq:prop}) given above.

\section{Applications and examples}
\label{ae}

In this section we analyze some simple consequences of the mirror
symmetry transformation that we have proposed. Apart from the case in
which only geometry and $B$-field are present, the situation will be
different from the
usual one for \cy's in that RR fluxes will transform, and so solutions
with some types of fluxes switched on get mapped generically to
solutions 
with other types of fluxes. On top of this we should also have the
usual exchange of K{\"a}hler and complex structure moduli, in the sense
of (\ref{eq:ejom}). Simple checks of both claims have been listed in
previous section; here we take these statements for granted and
examine the consequences.

The natural starting point is to check how the picture developed so far
reduces to known cases. We start from a brief discussion of an example
which has already been mentioned, and involves a CY manifold with
$B$-field turned on. This case was considered in \cite{glmw} in great
detail. Since the intrinsic torsion vanishes on CY, we start from $Q_{MN}$
built purely from components of $H$. The $Q_{m{\bar n}}$ gets a single
contribution from $H_1$.  If we follow \cite{glmw} and look for
a purely geometrical mirror, on the mirror side we may have non-zero
$\tilde{W}_3$ and $\tilde{W}_4 - \tilde{W}_5 =0$. Looking at $Q_{mn}$, we
see that the reality of remaining components of the flux ensures that on
the mirror side only $\tilde{W}^-_1$ and $\tilde{W}^-_2$ survive. This
agrees with \cite{glmw} up to a conventional $\pm$ exchange. So we recover
as a particular case the half-flat geometries and the G$_2$ lifts
discussed in \cite{hitchin2, cs}. Note that neither the starting
configuration, nor its mirror are vacua but rather domain walls.

The simplest background is when the $B$-field is turned off and we just
deal with \cy\ geometry. This case was also discussed in section \ref{sec:Mirror}, where
we recover the exchange of the complex structure and (the exponentiated)
K{\"a}hler form for mirror Calabi--Yau manifolds. An exchange of complex and
K{\"a}hler moduli for a metric of the form (\ref{eq:metric}) with $\lambda=0$ and the integrability properties of its complex structure were studied in \cite{lm}. Here we easily see that
the exchange of the $e^{iJ}$ and $\Omega$ is accompanied by an exchange of
their integrability conditions.

Without turning RR fields on, we can also consider yet another 
possibility of vacua.\footnote{Here and in the parts with also RR on,
the word vacuum should be understood with the usual grain of salt:
no--go theorems force us to consider noncompact or singular cases, or
to hope (in a less well--defined way) that some of the features
analyzed here will survive after taking into account higher--derivatives
corrections.} These cases are to obey the conditions 
given in (\ref{eq:hbg}). In our language these conditions read
$Q_{m n}=0=Q_{\bar m n}$. What one gets by the proposal (\ref{eq:prop})
is the condition $Q_{m \bar n}=0=Q_{\bar m \bar n}$, which is
obviously isomorphic to it (see the comments in previous section).
Less trivial is the statement that complex and K{\"a}hler moduli are
exchanged. To check this one would have first of course to know by
what groups moduli spaces are computed. 

This check we will not be able to perform here, and we limit ourselves
to some comments. First of all, in general moduli spaces of solutions
with fluxes are likely not to be simply factorized in K{\"a}hler and
complex part. This is because, unlike the \cy\ case, the conditions 
are no longer $dJ=0=d\Omega$, but something involving torsions; and
the definition of torsions (\ref{eq:djdo}) mixes $\Omega$ and $J$.
Also, in the \cy\ case the fact that the conditions were of simple
closure allowed to reduce the counting to a cohomology problem. In
general, here, we are dealing with conditions involving projections
 $P_{\mathrm{rep}} dJ$
and $P_{\mathrm{rep}}d\Omega$, where $P_{\mathrm{rep}}$ is a projector
on a certain representation. These conditions mean roughly that a form
is closed ``up to'' a contribution from the other form, schematically 
$dJ=\mathrm{operator}(\Omega)$. 
In general it should be possible to restate this as the cohomology of
a double complex. Coming back to
the case with $H$ only switched on, a preliminary analysis of moduli
spaces was sketched in \cite{beckersharpe}, following ideas in
\cite{rw}. 
Indeed the $H$--twisted
cohomology groups proposed there are total cohomologies of a double
complex with $\bar\de$ and $H^{2,1}\wedge$ as differentials.
We will unfortunately not say more on this here, but plan to come back
on the issue in the future. For now we just observe that, in known
examples, fluxes fix complex structure moduli. These considerations
tell us that in a mirror picture K{\"a}hler moduli will be fixed.

Type B solution for IIB strings presented in  \cite{mariana} provides with
another related case flux compactifications with the back-reaction taken
into account. The metric now is conformally CY, and
 RR-fluxes are turned on as well. In addition, supersymmetry conservation
imposes restrictions on  $H$-flux, which now turns out to be
primitive. We will not attempt here to present a complete analysis of the
mirror transformation and will  ignore the RR sector (which mirror
symmetry maps to RR fields in Type IIA theory). Thus our starting data
include $W_4 \sim W_5$ and $H_4$. Note that this means in particular that
we have $Q_{m \bar{n}} = 0$, and thus we need ${\tilde Q}_{mn} = 0$ on the 
mirror side. The two previous examples have this feature: we could
either take ${\tilde H} =0$ and $\tilde{W}_3=\tilde{W}_4-\tilde{W}_5 =0$
or have ${\tilde H}$ with imaginary selfdual primitive part and geometry
given by ${\tilde W}_3= *{\tilde H}_3$ and $2{\tilde W}_4= {\tilde
W}_5= 2d {\tilde \phi} =2i{\tilde H}_4$ as in \cite{strominger}. However
differently from previous cases $Q_{mn} \neq 0$ and this results in an
additional non-integrability of the complex structure on the mirror side
(in particular, ${\tilde W}^-_2$ cannot be zero now). Of course, explicit
constructions of such IIA string backgrounds would be of some interest.

The last application we will discuss here concerns the possibility of
lifting the SU(3) mirror symmetry picture to the G$_2$-structure case. We
could start from IIA string theory in a monopole background and lift it
to M-theory, using the explicit relations between the components of
intrinsic torsion for SU(3) and G$_2$ for U(1)-fibered manifolds. The
components of the torsion for the representations 1, 7 and 27 get complexified
by the corresponding representations of the $G_4$-flux. The analogy with
the SU(3) case is rather close. There as well there was a number of
components of the intrinsic torsion that get complexified by the $H$-flux;
mirror symmetry then mixed these with the components corresponding to
representations that are not contained in the flux (essentially $8 \oplus
8$ in that case, with some extra subtleties having to do with $3 \oplus
\bar{3}$ appearing twice). In the G$_2$ geometry, 14 is such a
representation, and the corresponding component of the torsion is the
lifting of $W^-_2$ \cite{cs,kmpt2}, 
the component of SU(3)-torsion central in the
exchange with the NS flux. Once more one would be hoping that going to
spinorial basis and writing for the invariant spinor the twisted 
covariant derivative 
will lead to a covariant expression for a mirror transformation
for the G$_2$ geometry. Indeed, as in (\ref{eq:deps}) the torsion for the
G$_2$-structure manifolds is also encoded in a covariant derivative $D_M
\epsilon = (q_M + i q_{MN} \gamma^N) \epsilon$, where $q$'s are real. Then
the eleven-dimensional supersymmetry transformations restricted to
seven-dimensions twist the covariant derivative by a term $\frac{i}{3} G_M
\epsilon - \frac{1}{3} (2G \delta_{MN} + G_{MN} + 2G_{NM})\gamma^N  
\epsilon,$ where we have defined $G \equiv \frac{1}{4!} G_{MNPQ}
(*\Phi)^{MNPQ}$,  $G_M \equiv \frac{1}{3!} G_{MNPQ} \Phi^{NPQ}$, 
$G_{MN} \equiv \frac{1}{3!} G_{MPQR} {(*\Phi)^{PQR}}_N$ using the
associative form $\Phi$. Putting all together we arrive at the twisted
operator
$$ 
D^G_M \epsilon = (Q_M + iQ_{MN} \gamma^N) \epsilon 
$$ 
which we can now use to extend the SU(3) mirror symmetry proposal.
Indeed, the G$_2$ analogue of (\ref{eq:prop})
can be written as \begin{equation}
  \label{eq:propG}
  Q^+_{[MN]} \longleftrightarrow - Q^-_{[MN]} \, \qquad Q_{\{MN\}}
\longrightarrow - Q_{\{MN\}} 
\end{equation}
where $\pm$ denote selfdual and antiselfdual representations respectively.
Note that only the former is complexified by $G$-flux, and
(\ref{eq:propG}) exchanges 14 with 7+7. In view of this, we may go back
to (\ref{eq:prop}) and note that there as well, modulo the trace part,
mirror symmetry can be thought of as an exchange of selfdual and
antiselfdual matrices ({\`a} la Hermitian Yang-Mills).

\section{Discussion}

We conclude by mentioning some open technical and conceptual problems.
Throughout the paper we have worked with a  $B_{\alpha \beta}=0$ case.
Obviously, this choice simplifies greatly the T--duality transformation. The
reason for this is most clear on the spinorial picture. As shown in
\cite{hassan},  the only change in the simple T--duality transformations 
used above (see section \ref{back}) occurs when $B_{\alpha \beta}$ component of
the $B$-field is nonzero. In this case we have to use instead 
$$
\psi_+ \to \psi_+ \ ,\qquad \psi_-\to e^{E} \gamma_f \psi_-
$$
where $E^{\alpha\beta}$ is defined in (\ref{eqn:invhB}).
We have here a gamma matrix exponential of $E\equiv
\frac12 E^{\alpha\beta}\gamma_{\alpha\beta}$ which has the
same form of the kappa--symmetry $\Gamma$ operator; in the power
series expansion  the products of all gamma matrices are antisymmetrized.

Note that without a $B_{\alpha \beta}$ component, there is a certain
ambiguity in the choice of T--duality invariants (\ref{eq:tinvs}). The
ambiguity is in the of  complexification by $H$ in $Q$'s. He have
chosen everywhere the plus sign (and correspondingly T--dual
expressions which become complex conjugates) for the following
reason. The singlet representation allows a simple calculation even
with a non-vanishing $B_{\alpha \beta}$. The result is then the first
formula in (\ref{eq:tw}), which fixes the ambiguity. For all other components we have chosen
the complexification rule consistent with that of $W_1$, hence the choice of
sign in the definition of the twisted covariant derivative
(\ref{eq:dheps}). The T--duality
rule for the spinors given above should allow to lift restrictions from
the $B$-field and verify this explicitly. We would like to emphasize
though that this restriction is of technical nature - for a number of
applications the $B$-field is generic enough. First, the $H$-flux
contains all the representations it can. Second, in the holomorphic
coordinate basis it is not hard to see that $B$ is of generic type and
contains both  $(1,1)$ and $(2,0)$ components. 
The latter is important for several aspects of topological B-branes 
(see \cite{kap} for a recent discussion, in which also Clifford(d,d)
spinors appear) and mirror symmetry \cite{kapor}.

Clearly there are two directions in which our results have to be
extended. As mentioned many times we have worked with a $T^3$
fibration with
finite-size fibers (and thus had a luxury of having extra vector
fields without zeros) and most of our formulae explicitly involve
restrictions to the base of the fibration. At the end we succeeded in
finding a basis in which the mirror/T--duality transformations can be
covariantized and written over the entire
six-dimensional manifold. The final simple rule for
the mirror transformation
$$
 Q_{mn} \longleftrightarrow - Q_{m\bar n}\ , \qquad Q_m \longleftrightarrow
  -{\bar Q}_m
$$
is of group-theoretical nature, and we conjectured it to be true for
general geometries, even without fibration structure at all, not even
locally. In particular it should also work when there is a fibration
but with singular fibers. From other side, singular $T^3$
fibers hold the key to SYZ picture, and would be extremely important
to understand their fate in any generalization of SYZ.

Finally, one would like to complete the picture by incorporating
D-branes. 
A better understanding of submanifolds in generalized CY
manifolds as well as vector bundles on these would be essential
preliminaries. Extending  the picture developed in \cite{lyz} for the
exchange of branes (a pair of calibration and bundle conditions) and
T--duality to generalized CY case would be of great interest. We may
recall once more that in SYZ picture both mirror manifolds appear as
moduli spaces for D-branes wrapping (sub)manifolds. One may hope that
eventually developing the picture of D-brane moduli spaces in
geometries with NS fluxes may lead to refining the proposal for mirror
symmetry presented here.

\bigskip\bigskip
{\bf Acknowledgments.} We would like to thank  Peter Kaste for
participation in early stages of this work, and Mariana Gra{\~n}a and Fawad
Hassan for useful conversations.
This work is supported in part by EU contract HPRN-CT-2000-00122 and by
INTAS contracts 55-1-590 and 00-0334.

\appendix
\section{Intrinsic torsion for $T^3$-fibered manifolds} \label{app:torsion}
The components of the intrinsic torsion are defined by
\begin{eqnarray*}
      dJ &=& -\frac{3}{2}\: {\rm Im}(W_1 \bar{\Omega}) + W_4 \wedge J + W_3, \\
 d\Omega &=& W_1 J^2 + W_2 \wedge J + \bar{W_5} \wedge \Omega.
\end{eqnarray*}
They can be computed using contractions ($\lrcorner$) with $J$ and $\Omega$:
\begin{eqnarray*}
  W_1 &=& \frac{4}{3}\, J^2 \lrcorner\, d\Omega
       =  -\frac{4i}{3}\, \Omega \lrcorner\, dJ
       =  \frac{1}{3}\, \epsilon_{ABC}\,
            (E^A \wedge E^B) \lrcorner\, dE^C\\
      &=& -\frac{i}{12}\, \epsilon^{ijk}\,
               V_{i\alpha}\, [d(V-i\lambda)]^\alpha_{jk} \\
\end{eqnarray*}
\begin{eqnarray*}
  W_2 &=& 4 J \lrcorner\, [ d\Omega - W_1 J^2 - \bar{W_5} \wedge \Omega ]\\
      &=& \frac{1}{12}\, \epsilon^{ijk}\,
           [ d(V-i\lambda) ]^\alpha_{jk}\;
           [ g_{pq} V_{i\alpha} - 3 g_{pi} V_{q\alpha} ] dz^p \wedge d\bar{z}^q \\
\end{eqnarray*}
\begin{eqnarray*}
  W_3 &=& dJ + \frac{3}{2}\, {\rm Im}(W_1 \bar{\Omega}) - W_4 \wedge J\\
      &=& \frac{3}{8}\, V_{i\alpha} [d\lambda^\alpha]_{jk}
                          \; dy^i \wedge dy^j \wedge dy^k\\
      & & -\frac{1}{4}\, [dV_\alpha]_{ik}
           [ \frac{3}{2} \delta_j^k \delta_\beta^\alpha
             + V_j^{\alpha} V^k_{\beta} - 2 V^{k\alpha} V_{j\beta} ]
                          \; dy^i \wedge dy^j \wedge e^\beta\\
      & & +\frac{1}{4}\, [d\lambda^\alpha]_{jk}
           [ \frac{1}{2} V_{i\alpha} V^j_{\beta} V^k_{\gamma}
            - \delta_i^j h_{\alpha\beta} V^k_{\gamma} ]
                          \; dy^i \wedge e^\beta \wedge e^\gamma\\
      & & -\frac{1}{8}\, V^i_{\beta} V^j_{\gamma} [dV_\alpha]_{ij}
                          \; e^\alpha \wedge e^\beta \wedge e^\gamma\\
      &=& \frac{1}{16} \left\{ -i[dV_\alpha]_{jk} V_i^\alpha + i[dV_\alpha]_{ij} V_k^\alpha
            + i[dV_\alpha]_{ki} V_j^\alpha
            + i[dV_\alpha]_{jl} V^{l\alpha} g_{ik} - i[dV_\alpha]_{kl} V^{l\alpha} g_{ij} \right.\\
      & & \phantom{\frac{1}{16}} \left. +i[d\lambda^\alpha]_{jk} V_{i\alpha}
            + i[d\lambda^\alpha]_{ij} V_{k\alpha} + i[d\lambda^\alpha]_{ki} V_{j\alpha} \right\}
          dz^i \wedge d\bar{z}^j \wedge d\bar{z}^k + \mathrm{c.~c.}\\
\end{eqnarray*}
\begin{eqnarray*}
  W_4 &=& 2 J \lrcorner\, dJ
       = \frac{1}{2}\, V^{\alpha k}\, [dV_\alpha]_{jk}\, dy^j\\
      &=& \frac{1}{2}\, h^{\alpha\beta}\,
           [ dh_{\alpha\beta} - {\cal L}_{gV_\alpha} V_\beta ]\\
\end{eqnarray*}
\begin{eqnarray*}
  W_5^{(1,0)} &=& - \Omega \lrcorner\, d\bar{\Omega}\\
      &=& \frac{1}{4}\left\{ V^k_\alpha\, [d(V+i\lambda)]^\alpha_{jk}
          + h^{\alpha\beta} \de_j h_{\alpha\beta} \right\} dz^j \\
      &=& \frac{1}{4}\, \left\{
           [ h_{\alpha\beta}\, {\cal L}_{gV^\alpha} V^\beta ]_j
           - i V^k_\alpha\, [ d\lambda ]^\alpha_{jk} \right\} dz^j \\
\end{eqnarray*}
where $dz^j = dy^j - i V^j_\gamma e^\gamma$.

In the last two expressions, we have used the Lie derivative ${\cal L}$, which is defined by
\begin{equation}
  {\cal L}_X Y = [X,Y] = [X^i \de_i Y^j - Y^i \de_i X^j] \de_j, \qquad
  {\cal L}_X \omega    = [X^i \de_i \omega_j + \omega_i \de_j X^i ] dy^j,
\end{equation}
on the vector field $Y$ and the 1-form $\omega$, with respect to the vector field $X$. We wrote $V^\beta$ and $V_\beta$ for the 1-forms $V_j^\beta dy^j$ and $V_{j\beta} dy^j$, while $gV^\alpha$ and $gV_\alpha$ are the vector fields $V^{i\alpha} \de_i$ and $V^i_\alpha \de_i$.\\

We also give here the components of the $H$ field
\begin{eqnarray}
  H &=& dB_2 \nonumber\\
    &=& \frac{1}{2}\, \partial_k B_{\alpha\beta} \,dy^k \wedge e^{\alpha} \wedge e^{\beta}
        + [\partial_k B_{i\alpha} - B_{\alpha\beta} \partial_k\lambda_i^\beta]
            \,dy^k \wedge dy^i \wedge e^{\alpha} \nonumber\\
    & & + \frac{1}{2}\, [\partial_k B_{ij} - \partial_k B_{i\alpha}\lambda_j^\alpha
                                           + B_{i\alpha} \partial_k \lambda_j^\alpha]
            \,dy^i \wedge dy^j \wedge dy^k
\end{eqnarray}
As a 3-form, we project $H$ on representations of SU(3) as we did for $dJ$:
\begin{equation}
  H = -\frac{3}{2}\, {\rm Im}(H_1 \bar{\Omega}) + H_4 \wedge J + H_3
\end{equation}
These components are computed with the same contractions used for $W$'s:
\begin{eqnarray}
  h_1 = H_1 &=& -\frac{4i}{3}\, \Omega \lrcorner\, H \nonumber\\
            &=& \frac{1}{12}\, \epsilon^{ijk} V_i^\alpha V_j^\beta \partial_k B_{\alpha\beta} \nonumber\\
            && +\frac{i}{12}\, \epsilon^{ijk} V_i^\alpha [dB_\alpha - B_{\alpha\beta} d\lambda^\beta]_{jk}
               \nonumber\\
            && -\frac{1}{12}\, \epsilon^{ijk} [\partial_k B_{ij}
               - \partial_k B_{i\alpha} \lambda_j^\alpha + B_{i\alpha} \partial_k \lambda_j^\alpha] \\
\nonumber\\
  H_4 &=& 2J \lrcorner\, H \nonumber\\
      &=& -\frac{1}{2}\, V^{k\alpha} [ dB_\alpha - B_{\alpha\beta} d\lambda^\beta ]_{jk} dy^j
          -\frac{1}{2}\, V^{k\alpha} \partial_k B_{\alpha\beta} \, e^{\beta} \\
      &=& h^4_k dz^k + \bar{h}^4_k d\bar{z}^k \nonumber\\
  h^4_k &=& \frac{1}{4} \left\{ V^{j\alpha} [dB_\alpha - B_{\alpha\beta} d\lambda^\beta]_{jk}
                            - i V^{j\alpha} \de_j B_{\alpha\beta} V_k^\beta \right\}
\end{eqnarray}
\begin{eqnarray}
  H_3 &=& H + \frac{3}{2}\, {\rm Im}(H_1 \bar{\Omega}) - H_4 \wedge J \nonumber\\
      &=& \frac{1}{4}\, V^{k\beta} V^{i\gamma}
            [\partial_k B_{i\alpha} - B_{\alpha\mu} \partial_k \lambda_i^\mu]
            \, e^{\alpha} \wedge e^{\beta} \wedge e^{\gamma} \nonumber\\
      & & +\frac{1}{2}\, \left[ \frac{5}{4}\, \partial_k B_{\alpha\beta}
                              - V^{j\gamma} V_{k\alpha} \partial_j B_{\gamma\beta}
                              - \frac{1}{2}\, V_k^\gamma V^j_\alpha \partial_j B_{\gamma\beta}
                         \right] dy^k \wedge e^{\alpha} \wedge e^{\beta} \nonumber\\
      & & +\frac{1}{8}\, [\partial_k B_{ij} - \partial_i B_{j\gamma} \lambda_k^\gamma
                                            - \partial_i \lambda_j^\gamma B_{k\gamma}]
           [V^k_\beta V^j_\alpha dy^i - V^k_\beta V^i_\alpha dy^j + V^i_\alpha V^j_\beta dy^k]
           \wedge e^{\alpha} \wedge e^{\beta} \nonumber\\
      & & +\frac{1}{4}\, [ (dB_\alpha)_{ik} - B_{\alpha\beta} (d\lambda^\beta)_{ik} ]
           \left[ \frac{3}{2}\, \delta^k_j \delta^\alpha_\gamma
                + V_j^\alpha V^k_\gamma - 2 V^{k\alpha} V_{j\gamma} \right]
           \, dy^i \wedge dy^j \wedge e^{\gamma} \nonumber\\
      & & +\frac{3}{8}\, [\partial_k B_{ij} - \partial_i B_{j\gamma} \lambda_k^\gamma
                                            - \partial_i \lambda_j^\gamma B_{k\gamma}]
           \, dy^i \wedge dy^j \wedge dy^k \nonumber\\
      & & -\frac{1}{8}\, V_j^\alpha V_k^\beta \partial_i B_{\alpha\beta}
            \, dy^i \wedge dy^j \wedge dy^k\\
      &=& \frac{1}{2}\, h^3_{ijk}\, dz^i \wedge d\bar{z}^j \wedge d\bar{z}^k + \mathrm{c.~c.}
          \nonumber
\end{eqnarray}
In analogy with $w$'s (see (\ref{eq:w2}) - (\ref{eq:w3t})) we have
introduced $h$:
\begin{eqnarray}
  h^3_{ij} &=& h^3_{ipq} {\Omega^{pq}}_j = h^3_{\{ij\}^0} + \frac{1}{3}\,h^3_t g_{ij} \nonumber\\
\nonumber\\
  h^3_{\{ij\}^0} &=& -\frac{1}{24}\,\epsilon^{pqk}\, [dB_\alpha]_{pq}
         [ 2 V_k^\alpha g_{ij} - 3 V_j^\alpha g_{ik} - 3 V_i^\alpha g_{jk} ] \\
  h^3_t &=& -\frac{1}{8}\,\epsilon^{pqk}\, [dB_\alpha]_{pq} V_k^\alpha \nonumber\\
        & & -\frac{i}{8}\,\epsilon^{pqk}\, [dB - \frac{1}{2}
              (dB_\alpha \wedge \lambda^\alpha + d\lambda^\alpha \wedge B_\alpha)]_{pqk}
\end{eqnarray}

\end{document}